\documentclass[]{spie}  %>>> use for US letter paper
\usepackage{amsmath,amsfonts,amssymb}
\usepackage{graphicx}
\usepackage[colorlinks=true, allcolors=blue]{hyperref}
\usepackage{caption}
\usepackage{subcaption}

\title{Holographic Beam Mapping of the CHIME Pathfinder Array}

\author[a,b]{Philippe~Berger}
\author[c,d]{Laura~B.~Newburgh}
\author[e]{Mandana~Amiri}
\author[f,g]{Kevin~Bandura}
\author[f]{Jean-Fran\c{c}ois~Cliche}
\author[a,c,d]{Liam~Connor}
\author[e]{Meiling~Deng}
\author[c,d]{Nolan~Denman}
\author[f]{Matt~Dobbs}
\author[e]{Mateus~Fandino}
\author[f]{Adam J.~Gilbert}
\author[e]{Deborah~Good}
\author[e]{Mark~Halpern}
\author[f]{David~Hanna}
\author[e,h]{Adam~D.~Hincks}
\author[e]{Gary~Hinshaw}
\author[e]{Carolin~H\"ofer}
\author[i]{Andre M.~Johnson}
\author[j]{Tom~L.~Landecker}
\author[e]{Kiyoshi W.~Masui}
\author[f]{Juan~Mena~Parra}
\author[a,c]{Niels~Oppermann}
\author[a,k,c,b,d]{Ue-Li~Pen}
\author[l]{Jeffrey~B.~Peterson}
\author[c]{Andre~Recnik}
\author[j]{Timothy~Robishaw}
\author[a,d]{J.~Richard~Shaw}
\author[f]{Seth~Siegel}
\author[e]{Kris~Sigurdson}
\author[m]{Kendrick~Smith}
\author[f]{Emilie~Storer}
\author[b,c]{Ian~Tretyakov}
\author[e]{Kwinten~Van Gassen}
\author[c,g]{Keith~Vanderlinde}
\author[e]{Donald~Wiebe}

\affil[a]{Canadian Institute for Theoretical Astrophysics, 60 St George St, Toronto, ON, M5S 3H8, Canada}
\affil[b]{Department of Physics, University of Toronto, 60 St George St, Toronto, ON, M5S 3H4, Canada}
\affil[c]{Dunlap Institute for Astronomy \& Astrophysics, University of Toronto, 50 St George St, Toronto, ON, M5S 3H4, Canada}
\affil[d]{Department of Astronomy \& Astrophysics, University of Toronto, 50 St George St, Toronto, ON, M5S 3H4, Canada}
\affil[e]{Department of Physics \& Astronomy, University of British Columbia, 6224 Agricultural Rd., Vancouver, V6T 1Z1, Canada}
\affil[f]{Department of Physics, McGill University, 3600 University St, Montreal, Canada}
\affil[g]{Lane Department of Computer Science and Electrical Engineering, West Virginia University, Morgantown, WV 26506, USA}
\affil[h]{Department of Physics, University of Rome `La Sapienza', Piazzale Aldo Moro 5, \newline I-00185 Rome, Italy}
\affil[i]{AMJ Consulting, West Kelowna, BC, V4T 1H7, Canada}
\affil[j]{National Research Council Canada, Dominion Radio Astrophysical Observatory, Box 248, Penticton, BC, V2A 6J9, Canada}
\affil[k]{Canadian Institute for Advanced Research, CIFAR Program in Gravitation and Cosmology, Toronto, ON, M5G 1Z8}
\affil[l]{McWilliams Center for Cosmology, Carnegie Mellon University, Department of Physics, 5000 Forbes Ave, Pittsburgh, PA, 15213, USA}
\affil[m]{Perimeter Institute for Theoretical Physics, 31 Caroline Street North, Waterloo, Ontario, \newline N2L 2Y5, Canada}

\authorinfo{Address inquiries to P. Berger. Email: pberger@cita.utoronto.ca}

\begin{document}

\maketitle
\vspace{-.1in}
\begin{abstract}
The Canadian Hydrogen Intensity Mapping Experiment (CHIME) Pathfinder radio telescope is currently surveying the northern hemisphere between 400 and 800 MHz. By mapping the large scale structure of neutral hydrogen through its redshifted 21 cm line emission between $z\sim0.8-2.5$ CHIME will contribute to our understanding of Dark Energy. Bright astrophysical foregrounds must be separated from the neutral hydrogen signal, a task which requires precise characterization of the polarized telescope beams. Using the DRAO John A. Galt 26 m telescope, we have developed a holography instrument and technique for mapping the CHIME Pathfinder beams. We report the status of the instrument and initial results of this effort.
\end{abstract}


\section{Introduction} \label{intro}

The Canadian Hydrogen Intensity Mapping Experiment (CHIME) is a new cylindrical transit interferometer currently being deployed at the Dominion Radio Astrophysical Observatory (DRAO) in Penticton, British Columbia. A smaller, two cylinder test-bed -- the CHIME Pathfinder -- has been built and instrumented with 128 dual polarisation dipole antennas and a custom FX correlator and is currently surveying the Northern hemisphere in 1024 frequency bands between 400 and 800 MHz. The Pathfinder correlator performs the full $N^2$ operation of correlating each of its 256 inputs at each frequency channel. See Ref. \citenum{chimepath1} for details of the design of the Pathfinder, Refs. \citenum{xeng1, xeng2, xeng3} for details on the GPU based X-engine, and Ref. \citenum{chimepath2} for a description of the calibration methodology.

As a transit interferometer, CHIME monitors the entire Northern sky visible from the DRAO each night. The telescope is optimized for 21 cm intensity mapping at redshifts $0.8-2.5$ where tomography of the large-scale distribution of neutral hydrogen (HI) will allow for a time-dependent measurement of the Baryon Acoustic Oscillations (BAO). The result will provide constraints on the time evolution of Dark Energy, including the epoch where it begins to dominate the energy density of the universe and so influences its expansion \cite{furl, moraleswyithe, bao1}. To do so, we must contend with astrophysical foregrounds, notably the synchrotron emission of the Milky Way, which are some five orders of magnitude brighter than the HI signal \cite{santoscoorayknox}. Removal of foregrounds is possible due to their smooth spectral nature versus the 21 cm signal, which should be relatively uncorrelated in frequency \cite{santoscoorayknox, shaw1, shaw2}. 

Foreground filtering is only possible with precise instrument characterization. Uncertainty in the primary beam leads to mode mixing, converting small-scale angular power into frequency structure. Uncertainty in the polarized response of the telescope leads to leakage of polarised signal into total intensity. Both of these effects can easily overwhelm the 21 cm signal. In Ref.~\citenum{shaw2}, these statements are made quantitative via fully polarized end-to-end simulations of a CHIME-like cylinder telescope. By varying the full width at half power of the illuminating dipole feed, the authors set the specification required for an unbiased estimate of the 21 cm power spectrum to $0.1\%$ of this parameter.

In these proceedings, we describe progress in mapping the full two-dimensional primary beam of each feed and frequency of the CHIME pathfinder array through a technique known as point-source holography \cite{radio1, radio2}. Holography is a well-known technique in radio astronomy and has been used with success in the near \cite{hol1} and far field \cite{hol2} to obtain high-resolution measurements on single-dish telescopes and dish arrays. Holographic techniques have further been used to map direction dependent polarisation leakages \cite{holpol}. In holography we track a bright point source with one telescope as a reference beam and correlate the signal with another telescope that is stationary. As the source transits, we measure a one-dimensional track through the stationary antenna beam. To serve as our tracking dish, we have equipped the John A. Galt 26 m telescope\cite{wolleben1, wolleben2} (hereafter 26 m), an equatorially mounted 26 meter diameter parabolic telescope also located at the DRAO, with a separate $400-800$ MHz receiver chain which is fed into the CHIME Pathfinder correlator. This allows correlation of its signal with the Pathfinder array. Since the Pathfinder is a fixed transit telescope, we observe sources at multiple declinations to obtain information on the North-South (NS) response of the beam. By averaging multiple transits from each source, we obtain high signal-to-noise, good angular resolution measurements of all 256 beams in both amplitude and phase. We have collected a preliminary data set, which allows for development and validation of the holographic analysis pipeline and initial results, that we present here.

The document is organized as follows. In Section \ref{sec1}, we describe the Pathfinder and 26 m instruments and the observations and data included here. In Section \ref{sec2}, we outline the holographic data analysis method. In Section \ref{sec3}, we discuss the processed results. Finally, in Section \ref{sec4} we discuss the results of full-sky simulations of the measurement which we have conducted to assess the effect of various systematics on the holographic reconstruction of the beam, notably the effect of background contamination.

\section{Instrument and Observations} \label{sec1}

\subsection{Instrument} \label{instrumentsec}

The CHIME Pathfinder\cite{chimepath1} consists of two $20\times37$ m$^2$ parabolic cylinders which are open North-South (NS) and focus incoming light East-West (EW). It is instrumented with 128 dual polarisation cloverleaf dipole feeds\cite{meiling}, 64 along the focal line of each cylinder, spaced by $0.3$ m. The illumination pattern of the two polarisations is known to be slightly elliptical so that the E-plane full width at half maximum (FWHM) of the NS polarisation is less wide than its H-plane, and vice versa for the EW polarisation. Therefore the FWHM of an EW slice of the NS polarisation is expected to be sharper than the EW polarisation. In general, we expect the main beam to be on the order of a degree in the EW direction, and to extend from horizon to horizon in the NS direction. The output of the Pathfinder correlator is a data cube of visibilities whose axes are frequency, cross-correlation, and time. In its nominal acquisition mode, it integrates from the sampling cadence of $2.56~\mu s$ to 20 s. However, in holography mode we sample at 10 s to ensure we well resolve the fringe rate of the Pathfinder-26 m baseline, which is about 150 m.

In error, the Pathfinder cylinders were aligned with Map Grid North which, at Penticton, is 1.9 degrees from celestial North. This alignment is accounted for in the analysis presented here. Another lesson learned from construction of the Pathfinder is that a poor choice of the form of the surface material lead to scalloping of the surface and medium-scale distortions which are larger than desired. In the following, particularly in Section \ref{fullarraysec}, we will see evidence that these distortions have an important impact on the resulting beam shape. Neither of these errors was repeated on the full-scale CHIME telescope.

For the measurements presented in this document we deployed custom $400-800$ MHz instrumentation at the focus of the 26 m telescope. The signal chain is depicted in Figure \ref{26mplot} (right) and consists of: a CHIME cloverleaf feed in a waveguide cavity with a flange and choke ring designed to symmetrize the polarized beams across the CHIME radio band; a pair of CHIME low noise amplifiers; an additional amplification stage at the focus; 100 m of RG214 cable; a set of ZFL-1000H+ Minicircuits amplifiers; 200 m of LMR400 cable; a final band pass filter amplifier stage; and finally the 26 m signal is digitized in the CHIME correlator.

\begin{figure}[h!] % not h only
	\centering
	\begin{subfigure}[a]{0.42\textwidth}
		\vspace{-4in}
		\includegraphics[width=\textwidth]{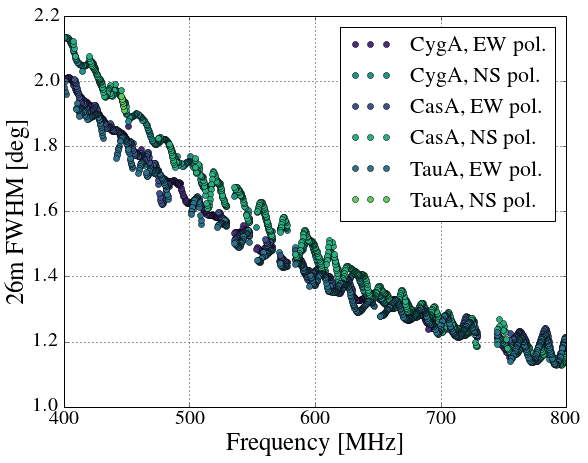}%
	\end{subfigure}
	\begin{subfigure}[b]{0.56\textwidth}%
		\includegraphics[width=\textwidth]{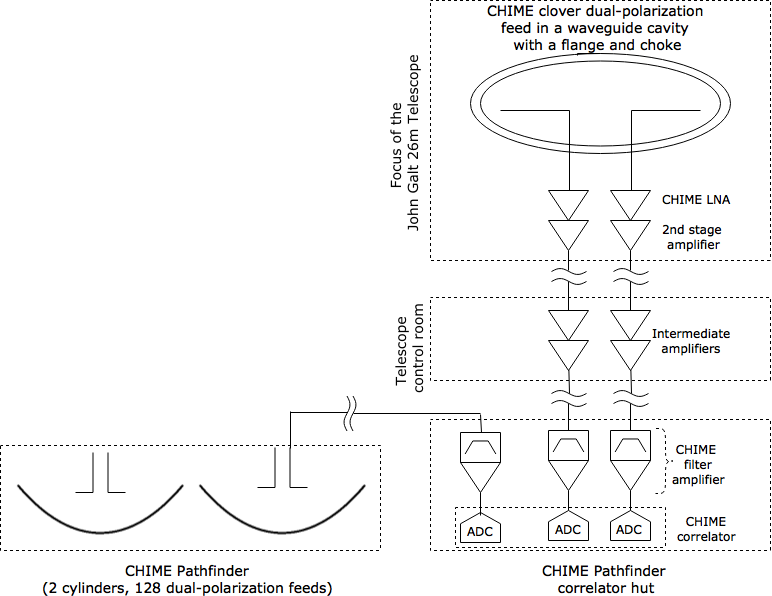}%
		\label{signalchain}
	\end{subfigure}%
	\caption{(Left) The FWHM of the 26 m beam as measured from drift scans centered on three different sources: Cygnus A, Cassiopeia A, and Taurus A. (Right) The analog signal chain of the 26 m instrument.}
	\label{26mplot}
\end{figure}

The telescope source equivalent flux density (SEFD) is expected to be $\sim$440 Jy, dominated by the first stage amplifier ($\sim$40 dB gain, $\sim$35 K noise temperature), and the instrument resolution should be $\sim1.5^{\circ}$ in the centre of the CHIME band. The beam and noise properties were measured with a set of drift scans and tracking scans of Virgo A, Cygnus A, Taurus A, and Casseiopiea A. A Gaussian profile was fit to the drift scans, and the resulting frequency-dependent FWHM values for both polarisations are shown in Figure \ref{26mplot} (left) for the drift scans of CygA, CasA, and TauA. The beam widths are in rough agreement with a simple model, but become slightly wider at low frequencies and differ more between polarisations at high frequencies than expected. In the single 1D slice from the drift scan, the FWHM of the two different polarisations are equal at high frequencies but differ at low frequencies by $\sim$6\%, compared to the expected 4\%. There is also a clear ripple in the FWHM, whose period is consistent with twice the path-length between the vertex and the focus of the dish (15.6 m, or 19.2 MHz). This standing-wave has been noted by other groups in the past on other wide band radio instruments \cite{wander1, wander2}.

Preliminary estimates of the noise temperature and resulting SEFD were calculated using various combinations of the above sources from both drifting and tracking scans, and are consistent with the expected value of 440Jy. However, there is a spread between measurements from different celestial sources of about 100Jy in the centre of the CHIME band, which will be the topic of further investigation.

\subsection{Observations} \label{obssec}

To measure the NS beam shape, we must use multiple point sources at different declinations. Table \ref{tab1} lists the sources chosen, as well as their positions and the number of observations used in the present analysis. The majority of these were conducted between mid February and early March 2016. However, a number of preliminary observations were made more than half a year earlier, in July 2015. The sources were chosen by ranking the brightest sources above the horizon of the Pathfinder, then selecting based on the need for night time observing. For this reason the Cygnus A observations considered here were taken from the preliminary observing phase. Observations start by positioning the 26 m on source a few hours before, and ended approximately symmetrically after, the transit of the source. This process is generally a competition between the requirement of observing extended sidelobe structure and maximizing the number of sources observed over the month-long period, which can lead to differing coverage and some asymmetry from night-to-night for a source.

\begin{table}[ht]
\caption{A list of the point sources chosen for the observations, along with the number of observations of each that were combined in the analysis.} 
\label{tab1}
\begin{center}
\begin{tabular}{| c  | c | c | c | c |  }
\hline
 Source  & RA    &        DEC    &  $N_{\rm obs}$ & S($\nu=681$ MHz) \\ \hline
Cygnus A & 299.88  & 40.73  & 2 & $3216 \pm 68$ \\ \hline
Taurus A & 83.62  & 22.03   & 4  & $1108.0 \pm 0.71$ \\ \hline
Virgo A & 187.71 & 12.39  & 4 & $399 \pm 20$\\ \hline
Hercules A & 252.80   & 4.99  & 2   & $105.8 \pm 2.2$  \\ \hline
Hydra A & 139.52  & -12.09  & 4 & $87.2 \pm 1.4$ \\ \hline
Perseus B  & 69.27  & 29.67  &  3  & $78.7 \pm 1.5$ \\ \hline
3C\_295 & 212.84  & 52.20 & 4   & $38.8 \pm 0.4$\\ \hline
\end{tabular}
\end{center}
\end{table}

\section{Data Analysis} \label{sec2}
The primary goal of this analysis is to use measurements of each observed radio source to recover measurements of the primary beam with good angular resolution in the EW direction, at the declination of the point source. In this section we describe the theory, data processing steps, and source information necessary to recover the final estimate of the two-dimensional beam. 

The CHIME Pathfinder correlator records the complex visibilities of a pair of sky channels. Following the notation of Ref.~\citenum{shaw1}, we denote the visibilities $V_{ij}$, or correlation between a channel $i$ and $j$ for an unpolarised sky as
\begin{align}
V_{ij}(\nu ; \phi) & \propto \int d^2\hat{n}~A_i(\hat{n}, \nu;\phi)A^*_j(\hat{n}, \nu;\phi) T(\hat{n}) e^{2\pi i \hat{n}\cdot\vec{u}_{ij}(\nu;\phi)},
\label{vis}
\\
\rightarrow V_{26,CH} & \propto A_{26}A^*_{CH}(\hat{n}_{\rm ps};\phi) T(\hat{n}_{\rm ps}) e^{2\pi i \hat{n}_{\rm ps}\cdot\vec{u}_{ij}(\phi)},
\label{holvis}
\end{align}
where $\nu$ is the frequency of observation, $\hat{n}$ is a direction on the sky, and $\phi$ is the celestial polar angle transiting the telescope at a given time of day. $T$ here can be understood as the brightness temperature of the sky although we have omitted the overall normalisation. $A_i$ denotes the primary beam of the $i$th channel. The $\phi$ dependence of $A_i$ highlights that, while the primary beam has some fixed two-dimensional shape, the pointing of the telescope rotates with the earth. Finally, $\vec{u}_{ij}=(\vec{b}_i - \vec{b}_j)/ \lambda$, where $\vec{b}_i$ are the positions vectors of the $i$th feed and $\lambda$ is the wavelength of observation. By definition, the $A_i$ are normalized to 1 on boresight. While the overall telescope response does indeed vary with frequency, this effect is degenerate with the gain of the amplifiers along the analog chain and is not the main focus of these proceedings, which are concerned with the $\hat{n}$ dependence of $A_i$ at each frequency. 

In holography mode, the voltage response of a dish telescope (in our case the 26 m) tracking a bright point source is correlated with that of the transit telescope (in our case the CHIME pathfinder). The integral in Eq. (\ref{vis}) collapses since the point source is expressed as a delta function at its sky location $\hat{n}_{\rm ps}$, giving Eq. (\ref{holvis}) (where we have dropped the explicit frequency dependence). Furthermore, the effect of the tracking dish is constant and the only remaining time dependence, aside from the geometric phase, is the primary beam of the CHIME pathfinder. The visibilities from a holographic transit yield the shape of the Pathfinder primary beam at the declination of the point source.

\begin{figure}[h!]
\centering
\includegraphics[width=\textwidth]{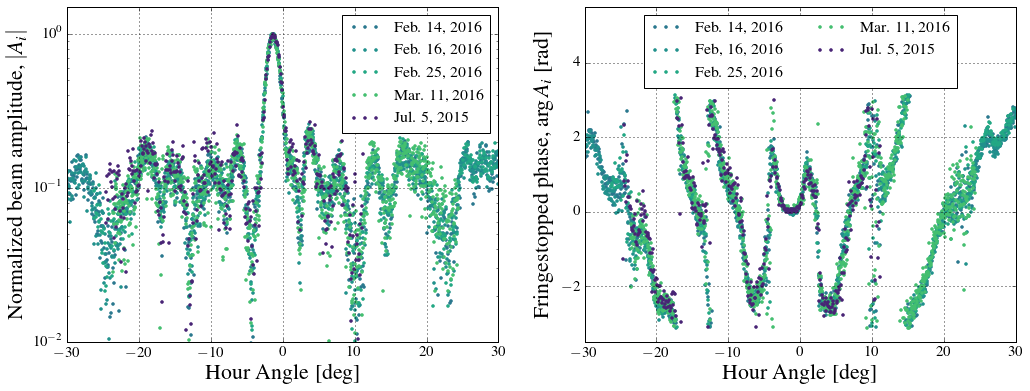}
\caption{Visibilities of a CHIME-26 m correlation channel for a set of holographic scans of Virgo A, for a single East-West polarisation channel at 681 MHz. Amplitude and phase for the four observations included in the analysis are compared to data from preliminary observations from July 2015. The amplitude has been normalized to one at transit, while the phase has been fringestopped (which normalizes to zero at transit).}
\label{virraw}
\end{figure}

Figure \ref{virraw} displays the raw holographic visibilities from the five Virgo A transits included this analysis, for a single East-West (EW) polarisation channel on the eastern Pathfinder cylinder, in amplitude (left) and phase (right). We recover high signal-to-noise measurements of both the main beam and far sidelobe structure. The results are seen to be reproducible over a period of months. The data are uncorrected except for normalisation and removal of the geometric phase in Eq. (\ref{holvis}), known as fringestopping. Before fringestopping, the Fourier components of the holographic visibilities occupy a band in Fourier space whose width corresponds to the physical EW width of the Pathfinder cylinder but whose centre corresponds to the Pathfinder-26 m baseline. Fringestopping removes the phase associated to the baseline, which symmetrizes the Fourier components about zero. This establishes the Fourier correspondence between the measured beam pattern and the aperture illumination (discussed at length in Ref. \citenum{radio2}, for example).

\subsection{Flagging, pre-filtering, and fringestopping}
From the visibility data cube with frequency, baseline, and time axes, we first select the elements corresponding to correlations between the 26 m and Pathfinder channels. Data shown here is such that the feed pairs between CHIME and the 26 m are oriented parallel at transit. Then, a simple flagging for radio frequency interference (RFI) is performed by a median absolute deviation thresholding along the time axis. Next, we perform a delicate high pass filter along the time axis, removing only the lowest frequency modes which are unphysical and can only correspond time independent or slowly varying gains. This is designed to mitigate the effect of gain fluctuations and to protect the phase from spurious signals that would be shifted during fringestopping. This does not affect the peak height of the amplitude at transit. We can then fringestop and normalize the phase to be zero at transit. We do not normalize the amplitudes because the relative peak heights of the various sources will be used to obtain the NS shape of the beam.

\begin{figure}[h!] % not h only
	\centering
	\begin{subfigure}[b]{0.49\textwidth}%
		\includegraphics[width=\textwidth]{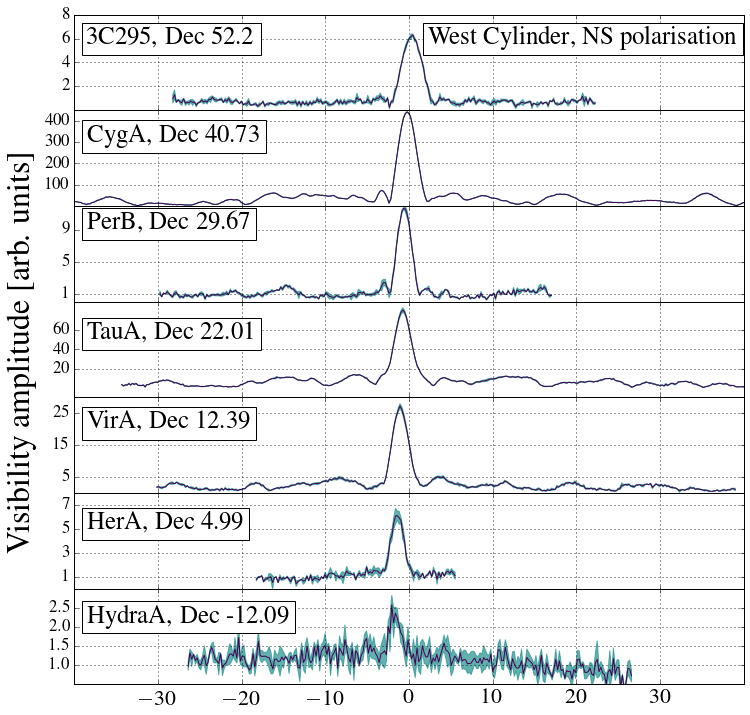}%
	\end{subfigure}%
	\begin{subfigure}[b]{0.47\textwidth}
		\includegraphics[width=\textwidth]{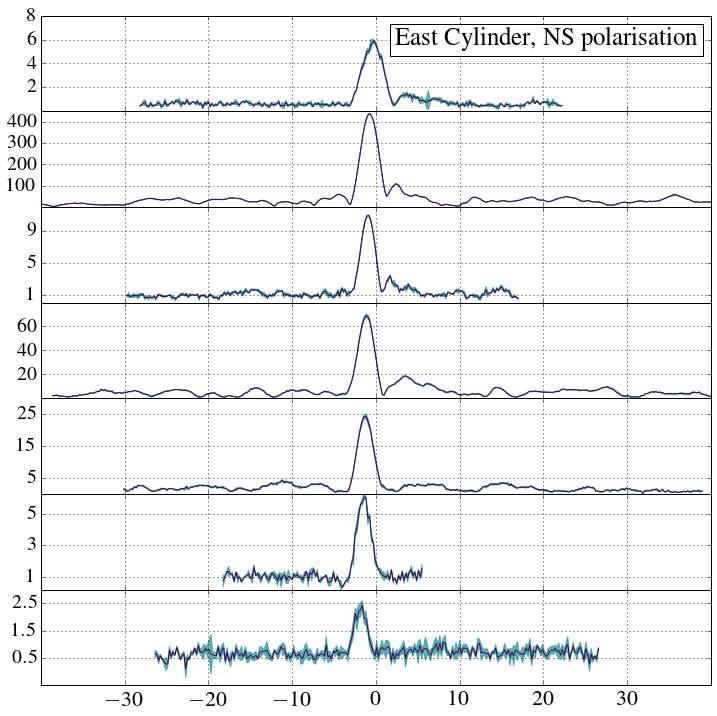}%
	\end{subfigure}
	\begin{subfigure}[b]{0.4925\textwidth}
		\includegraphics[width=\textwidth]{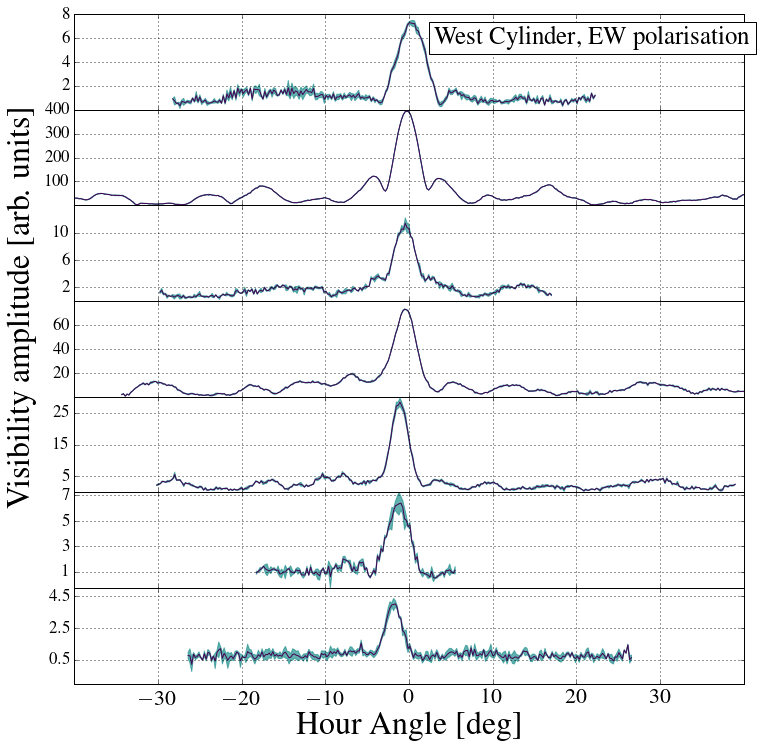}%
	\end{subfigure}
	\begin{subfigure}[b]{0.4675\textwidth}
		\includegraphics[width=\textwidth]{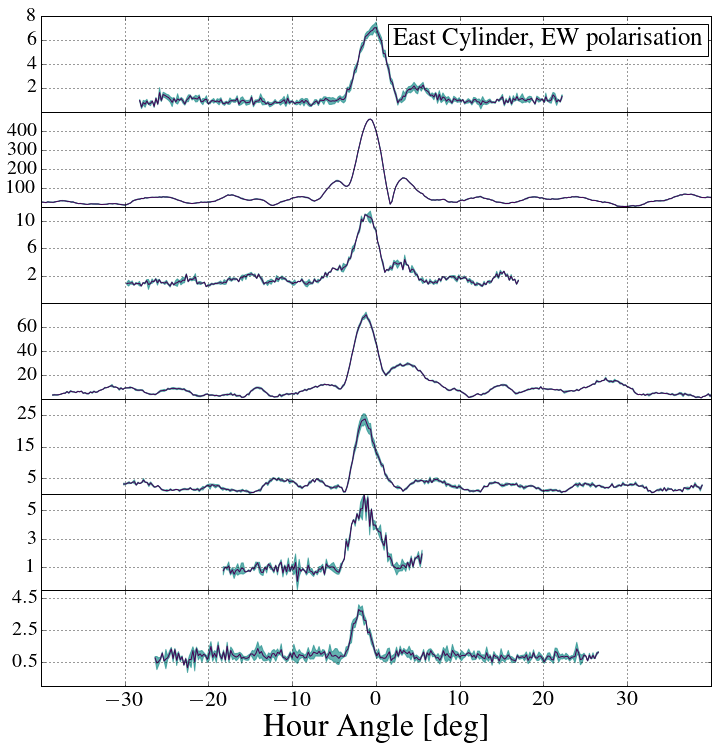}%
	\end{subfigure}
\caption{The amplitudes of the gridded and averaged visibilities for all sources at 681 MHz for both polarisations of a single antenna near the centre of the West cylinder, and its corresponding antenna on the East cylinder. Regions of one standard deviation are shaded. The offset of the peak of the main beam is due to a 1.9 degree rotation of the Pathfinder cylinder axis from astronomical North.}
\vspace{-.1in}
\label{griddeddata}
\end{figure}

\subsection{Gridding and averaging}
We first regrid the visibilities in hour angle, using an inverse Lanczos resampling.\footnote{https://en.wikipedia.org/wiki/Lanczos\_resampling} At this point the only input to the noise covariance other than the assumption of constant instrumental noise across frequencies and baselines is RFI flagging, which assigns infinite noise to flagged time samples. We note that the telescope is not sensitive to spatial Fourier modes larger than its width ${\rm w_{EW}}$ in wavelengths
\begin{equation}
m_{\rm max} = 2\pi \frac{\rm w_{EW}}{\lambda},
\end{equation}
where $m$ indexes the Fourier mode conjugate to the azimuthal angle $\phi$. Therefore, we conservatively choose to bin on the angular scale corresponding to $4\times m_{\max}$ at each frequency to ensure these modes are well-sampled and to prevent any binning artifacts. While further binning would produce smoother results, we prefer to leave this work to the averaging procedure. 

Next we produce an estimate of the peak response at each feed and frequency by fitting a Gaussian to the amplitude and averaging the peak value across observations. This also provides an estimate of the relative uncertainty between the transits in the covariance of the Gaussian parameters. To produce the final beam shape, we scale each observation to the average peak value and then average separately for real and imaginary parts, assuming the noise is uncorrelated between observations. We include only sections of beam for which we have collected at least 2 transits. The differing coverage means the final estimate can have different noise properties for different sections. An estimate of the final errors $\sigma_{|A|}$ and $\sigma_{\arg{A}}$ is provided simply by the standard deviation of the non-zero contributions to the averaging.

\begin{figure}[h!] % not h only
	\centering
	\includegraphics[width=0.6\textwidth]{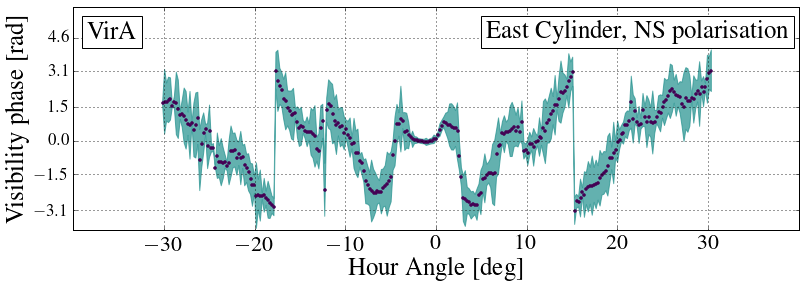}%
	\caption{The phase of the gridded and averaged beams at 681 MHz for a selected antenna and polarisation for Virgo A. Shaded regions represent regions of one standard deviation of the averaged observations $\sigma_{\arg{A}}$.}
	\label{phaseav}
\end{figure} 

Figure \ref{griddeddata} displays the amplitudes of the multi-declination results at 681 MHz for both polarisations of a single antenna near the centre of the West cylinder, and its corresponding antenna on the East cylinder. Regions of one standard deviation are shaded. The units of the amplitudes here are arbitrary as no absolute calibration has been attempted. These values are meaningful relative to each other as the product of the amplitude of the primary beam at the declination of the source and the flux of the source (Eq. (\ref{holvis})). Figure \ref{phaseav} shows the phase recovered from the averaging procedure for Virgo A at a single frequency, feed, and polarisation.

It is clear that many more transits of lowest signal-to-noise sources will be necessary to reach the accuracy obtained from even a single Cygnus A observation. Currently, they provide a first estimate of the NS dependence of the main beam, while the structure at large angles is still noise dominated. We do observe some sidelobe structure which is similar in the NS direction between sources spanning the range of S/N, and much of the far sidelobe structure is common between high S/N sources. The changing offset from zero hour angle of the peak of the main beam with declination is a known effect, due to the 1.9 degree rotation of the Pathfinder cylinder axis from astronomical North, discussed in Section \ref{sec1}. 

\subsection{Source flux renormalisation for 2D beam estimation}

To produce the final estimate of the amplitude of the 2D beam at each feed we must divide out the dependence on the flux of the source, corresponding to the $T(\hat{n}_{ps})$ dependence in Eq. (\ref{holvis}). Since we are only interested in the relative scaling of the main beam between the sources, we normalize each transit so that the peak value of a Gaussian fit to the Cygnus A track is 1. To do so we require an estimate of the relative fluxes of the point sources in the CHIME band. We combine data from Baars {\it et al.} 1977 \cite{baars}, NVSS \cite{NVSS}, VLSS \cite{VLSS}, 3CRR \cite{3CRR}, and WENSS \cite{WENSS} and perform parametric fits to the spectra of our sources. We adopt a three-parameter model for the spectral flux densities of our sources. We estimate the spectral indices from the data by a maximum likelihood method, incorporating the reported uncertainties on each measurement. The estimated fluxes at $681$ MHz are provided in Table \ref{tab1}. With the derived spectra in hand, the beam amplitude at the declination of the source relative to Cygnus A is calculated by dividing by the ratio of the flux of the source to that of Cygnus A and by the original peak value of the Cygnus A track. In all steps the errors are assumed to be Gaussian and uncorrelated and are therefore propagated using standard methods.

\section{Results and Discussion} \label{sec3}

In this section we display the final estimate for the two-dimensional Pathfinder beam as measured from the holographic data set. We also discuss some of the basic properties of the beam, in comparison to the fiducial theoretical model and to numerical simulations of an ideal Pathfinder cylinder. In general, the primary beam slices we recover meet our expectation from our 7 sources, in terms of main beam shape and S/N on sidelobe structure. As well, the measured amplitude of each source will broadly show the beam shape dropping off away from the zenith in the NS direction. This is the first measurement of CHIME Pathfinder beams, and shows the promise of the holographic technique for achieving the required calibration requirements (discussed in Section \ref{intro}). We will describe future directions in improving this technique.

\subsection{Single slice properties}

\begin{figure}[ht] % not h only
	\centering
	\includegraphics[width=0.95\textwidth]{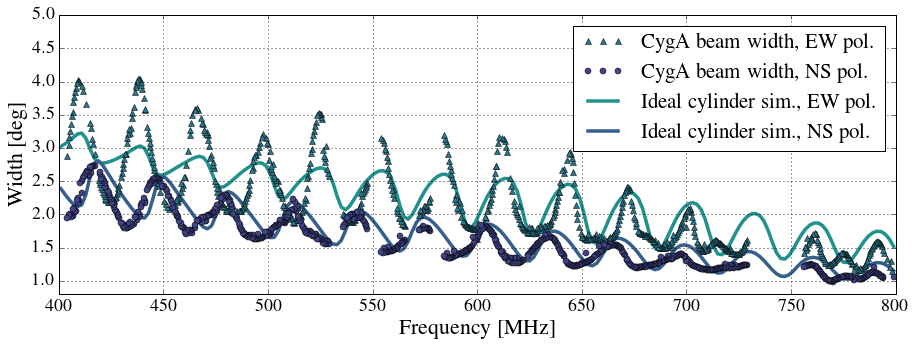}%
	\caption{The beam width of the Cygnus A slice as a function of frequency, scaled to account for the declination of the source, for both polarisations of a single feed on the East cylinder. Also plotted are the predictions from numerical simulations of an ideal parabolic cylinder illuminated by a CHIME Pathfinder feed. The dominant mode of the oscillation has a period of $\sim$ 30 MHz, matching the light travel time of twice the 5 m focal length of the Pathfinder. This is a well-known phenomenon in on-axis telescopes \cite{wander1, wander2}. See Figure \ref{26mplot} and also Ref. \citenum{wolleben1} for a similar example on the John A. Galt 26m telescope.}
	\label{beamwidths}
\end{figure}

The fiducial theoretical model for the CHIME beam, derived in detail in Ref.~\citenum{shaw2}, consists of an ideal parabolic cylinder and dipole at the focus. There are two free parameters, the $E$-plane and $H$-plane FWHM of the dipole, which control the illumination across and along the cylinder. The resulting primary beam is the product of a focused shape in the EW direction, and a long extended shape in the NS direction. This model has maximal symmetry in the telescope coordinate system (whose origin in latitude and longitude is at zenith), which we denote with $(\theta_t, \phi_t)$. In this simple model, the incoming plane wave that reflects only once off of the cylinder to the feed and so predicts a beam FWHM that increases monotonically with wavelength, and whose $E$ and $H$-plane widths determine both the resulting beam width and sidelobe level. These predictions, including the symmetry of the beam, can be broken in the realistic case of a distorted cylinder and interference effects caused by multiple bounces. 

\begin{figure}[htp] % not h only
	\centering
	\includegraphics[width=0.92\textwidth]{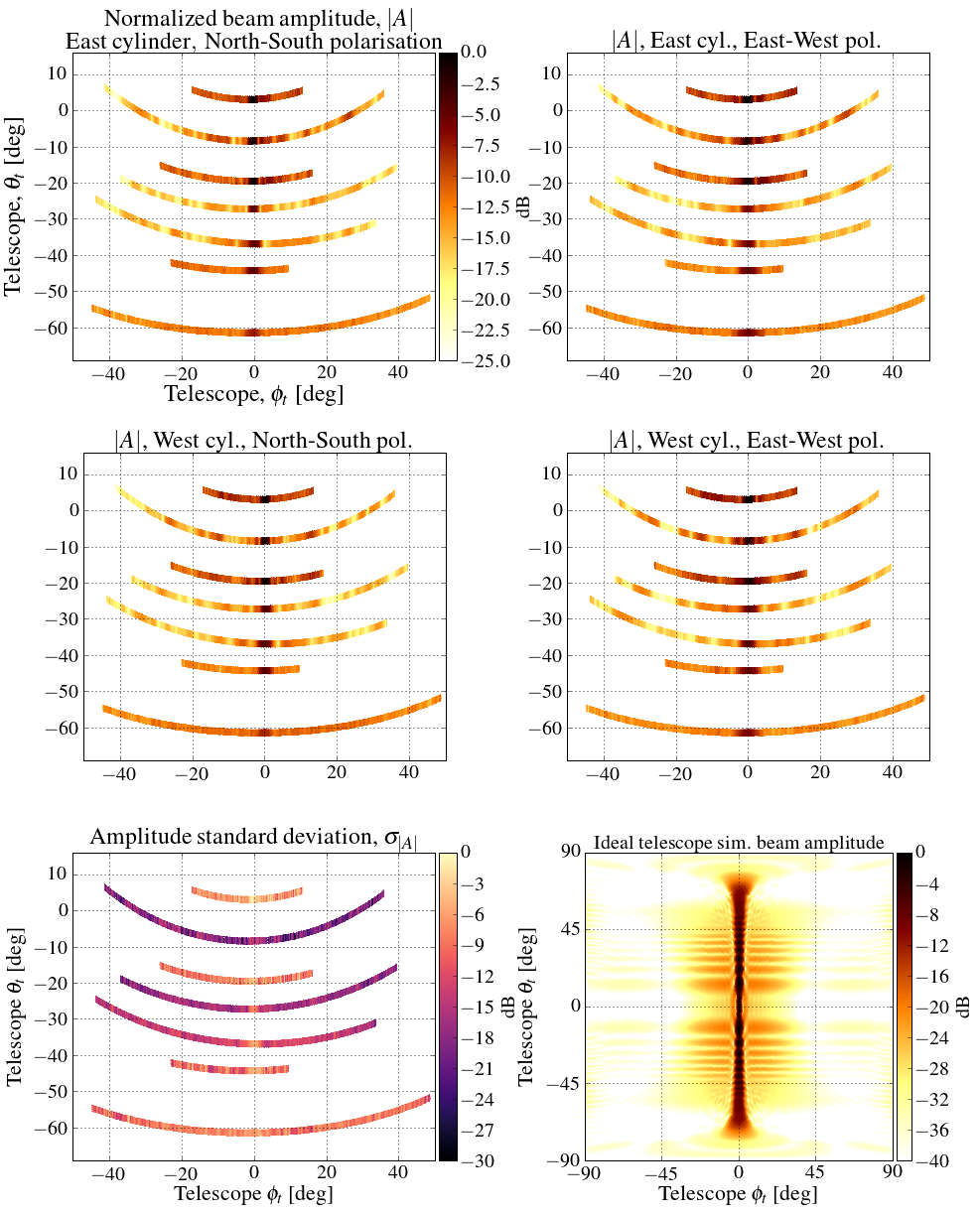}%
	\caption{The top four panels: A cartesian projection of the 2D beam estimate for all cylinders and polarisations, in latitude and longitude of telescope coordinates, for a sample feed at 681 MHz. The traces have been artificially thickened, which may misrepresent localized features. These are slices through a full pattern like the model shown in the lower right hand panel. Bottom left: The standard deviations, which vary by several orders of magnitude based on the signal-to-noise of the corresponding source. Bottom right: The full 2D beam amplitude from a simulation of an ideal cylinder at 681 MHz. In all cases the maps have been normalized to the peak value.}
	\label{full2dbeam}
\end{figure}

It is well-known\cite{wander1, wander2} that the presence of a standing wave between the focal line and reflective surface causes oscillations in the FWHM and forward gain with frequency. Figure \ref{beamwidths} shows the FWHM of the Cygnus A beam slices as a function of frequency, for both polarisations. Also plotted are the beam widths from electromagnetic numerical simulations of an ideal parabolic cylinder. These simulations include no blockage or distortion effects, but include a model of the CHIME Pathfinder feed\cite{meiling} and interference from a second bounce (see also Figure \ref{full2dbeam}). The FWHM displays a ripple in frequency with a period $\sim$30 MHz, consistent with a standing wave between the focal line and reflective surface. The differing median width between the two polarisations is predicted by the simple differing illumination included in the theoretical model (see Section \ref{instrumentsec}) and our numerical simulations. However we find that the amplitude of the ripple is larger than the simulations would suggest. We also find variations in the telescope pointing with frequency which correspond to the width oscillations. We have run simulations which include both distortions in the cylinder surface and multiple reflective bounces and see similar effects in the results. We are currently investigating this relationship with field tests and further simulations.

We observe clear sidelobe structure which is common between sources at nearby declinations, however the structure can vary with cylinder and polarisation. Since the two polarizations illuminate the cylinder differently, it is not unexpected that the two polarizations may have slightly different sidelobe structure. Similarly, the two cylinders are not identical, and so the sidelobes should be different between them. Again, we see similar broken EW symmetry effects in simulations of a cylinder with an imperfect surface which is itself asymmetric. Furthermore, surface distortion amplifies sidelobes uniformly, to a level which matches that which we observe in the data.

\subsection{2D beam properties}

Figure \ref{full2dbeam} displays the amplitudes of the traces obtained from the averaging procedure of section \ref{sec2}, however scaled by the fluxes of the sources to reflect the shape of the NS beam, and transformed from celestial to telescope coordinates. We show the results for a sample parallel set of feeds on each cylinder, for both polarisations, and for a single frequency bin at 681 MHz. The bottom left panel shows the standard deviations for the top left panel, which are representative for the full set. There are a few orders of magnitude between the S/N of our brightest three sources and the dimmest four. Clearly we are still noise dominated at large angles in the latter four. As remarked in the last subsection, we observe correlated sidelobe structure between the brightest sources in the NS direction. Also, there is a large asymmetric sidelobe at small positive $\phi_t$ which correlates between all sources. The NS correlation is only approximate, but this is to be expected since the incoming plane wave strikes the cylinder at different angles, and so probes different sections, depending on the source. The traces in Figure \ref{full2dbeam} are artificially thickened and so may also misrepresent localized features.

Even the low S/N sources allow us to infer the shape of the main beam with declination. We observe a plateau which extends to about $20^{\circ}$ from zenith, followed by a sharp drop off. This is in accordance with the prediction from our ideal cylinder simulations, discussed in the last subsection. We plot the full 2D beam shape from these simulations for reference in the bottom right panel of Figure \ref{full2dbeam}, at 681 MHz. It is clear, from comparison with simulation, that there is much more structure in the NS direction than we have sampled with only 7 sources. We leave the discussion of our plans for future work in this area to the conclusion, Section \ref{conclusion}.

\subsection{Full array, redundancy} \label{fullarraysec}

\begin{figure}[ht] % not h only
	\centering
	\begin{subfigure}[b]{0.49\textwidth}
		\includegraphics[width=\textwidth]{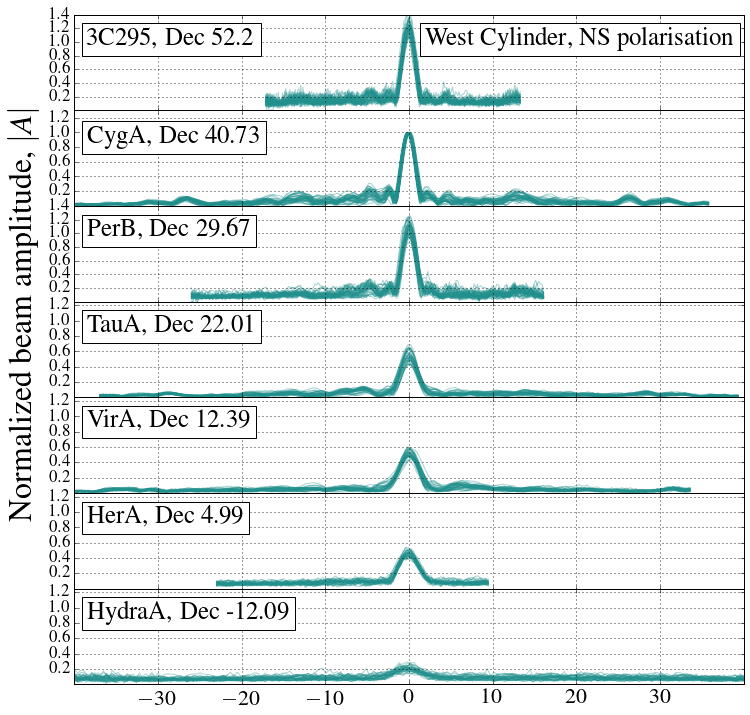}%
	\end{subfigure}
	\begin{subfigure}[b]{0.465\textwidth}
		\includegraphics[width=\textwidth]{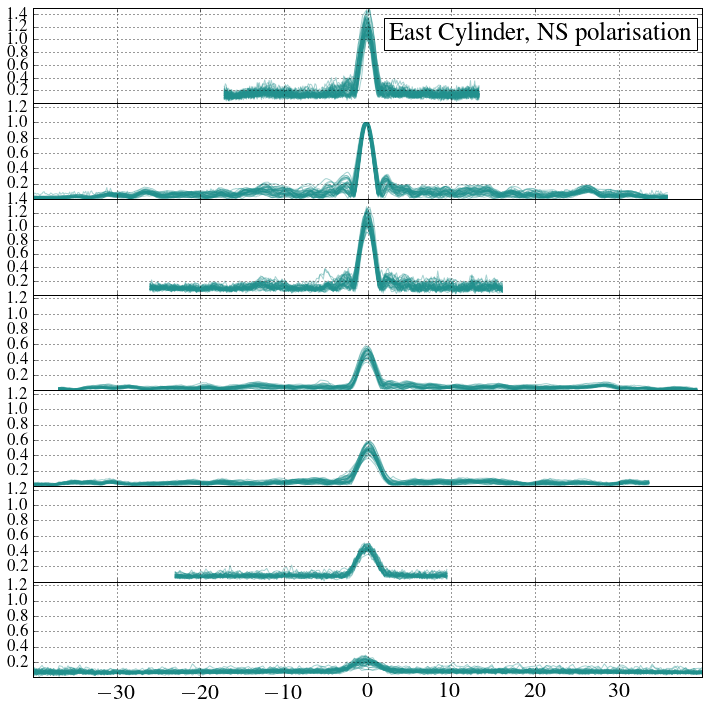}%
	\end{subfigure}
	\begin{subfigure}[b]{0.49\textwidth}
		\includegraphics[width=\textwidth]{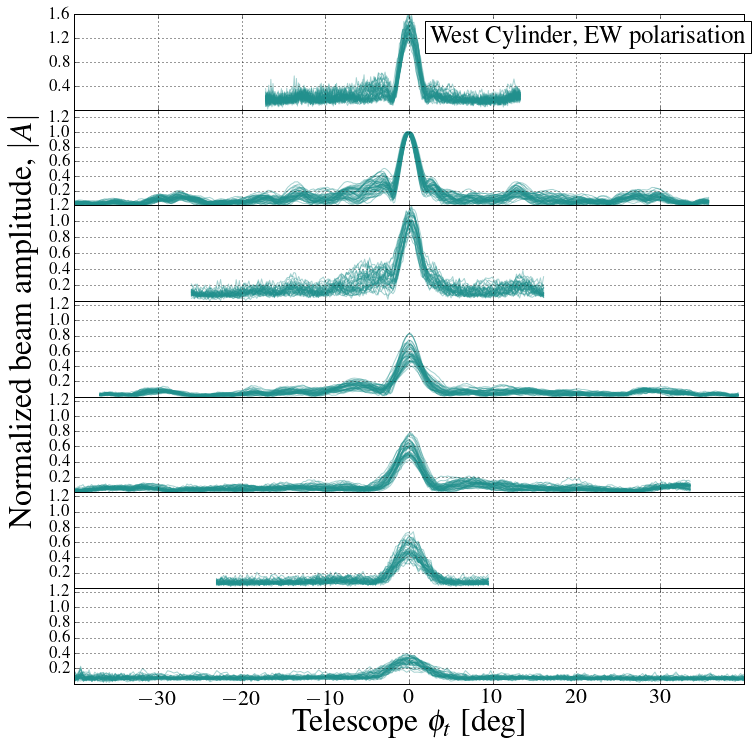}%
	\end{subfigure}
	\begin{subfigure}[b]{0.46\textwidth}
		\includegraphics[width=\textwidth]{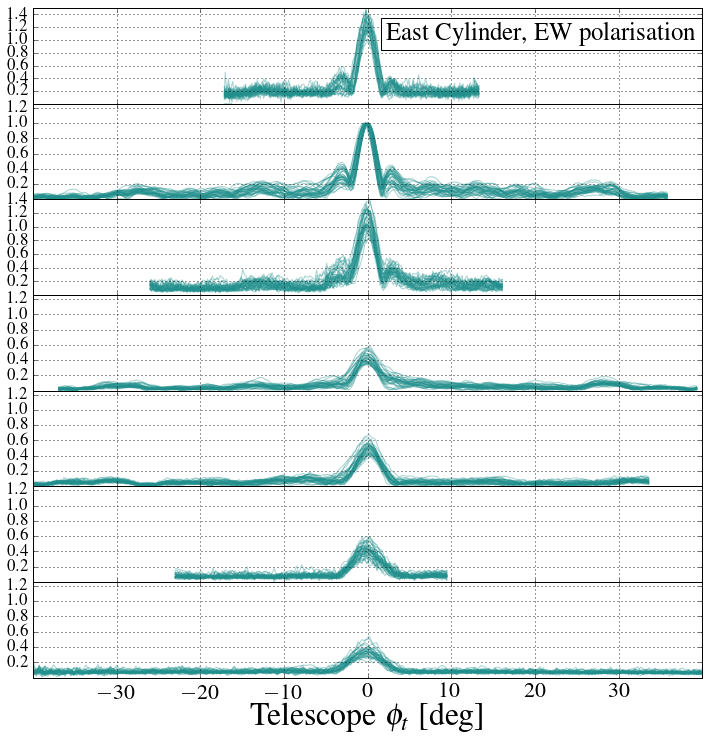}%
	\end{subfigure}
\caption{The averaged amplitude data for all feeds at 681 MHz are over plotted, normalized to the peak of the CygA slice and corrected for coordinates so to trace the shape of the 2D beam. }
\label{allfeeds}
\end{figure}

Figure \ref{allfeeds} shows the averaged amplitude of the 2D beam, one line for each feed of the array, for a single frequency bin at 681 MHz. The data have now been scaled by the fluxes of the sources (and normalized to Cygnus A) so that the NS beam shape can be inferred, and transformed to telescope coordinates. Much of the of sidelobe structure, even at large angles, is highly correlated across full sets, by cylinder and polarisation. Of course there is some variation within each panel of Figure \ref{allfeeds}, owing to the feeds probing different areas along the cylinder. This spread is larger for the EW polarisation since, as discussed in the previous subsection and Section \ref{instrumentsec}, the ellipticity of the illumination means the EW polarisation sees more of the cylinder.

An advantage of a phased array such as the CHIME pathfinder is the high redundancy of its baselines, due to the placing of the feeds on a regularly spaced grid. This choice, while not ideal for imaging, provides maximal sensitivity for a selected number of spatial Fourier modes. As well, it can greatly simplify the task of calibrating such a large number of antennas \cite{redundantbaselines}, as each redundant baseline should see the same sky. Unfortunately, variations in the primary beam from antenna to antenna break the redundancy to some degree and so complicate this analysis.

One of the biggest advantages of our holographic set up is that a single scan correlates the 26 m reference beam with the entire CHIME Pathfinder array, and so provides high signal-to-noise ratio measurements of each feed individually, at each frequency. Knowledge of the difference between feeds can be used produce an accurate estimate of the true sky modes, restoring the redundancy of the array. 

\section{Simulations} \label{sec4}

In this section, we describe the results of simulations conducted to investigate the effect, if any, of various systematics on the holographic signal we measure. In the previous sections, we applied a holography technique to measure the primary beam of a cylindrical transit telescope, which covers a $\sim 1^{\circ}$ strip of the sky and displays non-negligible sidelobe structure. Two primary concerns in evaluating whether the sidelobe structure we observe is accurate is whether it could be contamination from the galaxy, nearby point sources, or effects from resolving the source itself. Eq. (\ref{holvis}) could be a bad approximation to Eq. (\ref{vis}), whose integral extends over the entire sky and so could include contributions other than the point source, or the point source. The primary beam of the tracking dish will have its own sidelobes, which could allow sources at large angles to the primary point source to correlate. Alternatively, we realize that the 26 m-Pathfinder baseline is on the order of 150 m which means that the angular extent of some of our sources could begin to cause frequency dependent effects in the recovered holographic beam. 

Our simulation code uses the publicly available radio cosmology codes \texttt{cora} and \texttt{driftscan}\footnote{https://github.com/radiocosmology/}, which allow for simulation of a fully polarised radio sky and observation thereof by an interferometer in drift scan mode. The problem, for a Pathfinder-like array, is rendered computationally tractable by exploiting the $m$-mode formalism \cite{shaw1, shaw2}. We have written an extension to \texttt{driftscan} that allows for simulation of a drifting array in holography mode with a tracking paraboloidal dish. The \texttt{cora} foreground model, described in detail in Ref~\citenum{shaw2}, contains a catalogue of known point sources, as well as a Gaussian background of unresolved sources and Galactic model whose intensities are based on Ref.~\citenum{santoscoorayknox}. The model is then augmented with polarised point source and Galactic emission, and Faraday rotation. In order to properly resolve the scales of a realistic 26 m-Pathfinder baseline we perform our simulations with a \texttt{healpix}\footnote{http://healpix.jpl.nasa.gov/}\cite{gorski} \texttt{nside} of 1024.

\subsection{Extended sources}

Inspecting our list of sources there are two examples, Virgo A and Taurus A, which have angular extent on or near the resolution of our longest 26 m-Pathfinder baseline at 800 Mhz, of about $7'$. 
\begin{figure}[h] % not h only
	\centering	
	\includegraphics[width=0.95\textwidth]{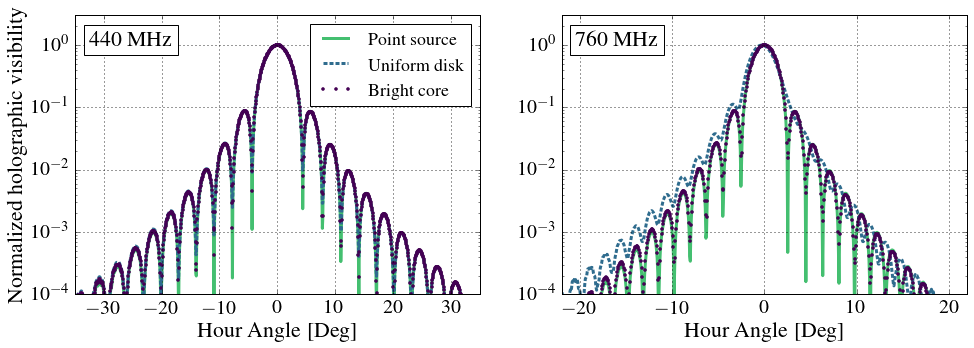}%
	\caption{The amplitudes of the simulated Taurus A beam slices for three flux distribution models of increasing realism, at 440 and 760 MHz. Only in the unrealistic model of an ellipsoid of uniform surface brightness do we observe frequency dependent effects in the recovered beam.}
	\label{crabsim}
\end{figure}

Taurus A is a $\sim 4.5'\times 7'$ (at the limit of our maximum angular resolution) ellipsoid whose emission is quite spread across its surface area \cite{crab}. We simulate three cases: the ideal point source, one where the flux is evenly spread over the entire ellipsoid, and the most realistic model where a dense core of $30\%$ of the surface area contains $50\%$ of flux. We also tilt the ellipse at $45^{\circ}$. Figure \ref{crabsim} displays the results of our simulations for the three Taurus A models. Only in the unrealistic case of uniform surface brightness do we see frequency dependent effects in the holographic slices.

Virgo A consists of a bright core containing about 90$\%$ of the flux and a diffuse halo extending to $14'$ \cite{baars, m87lofar}. We simulate this by comparing a sky which contains only a single non-zero pixel containing all of the flux of Virgo A, to a single pixel with 90$\%$ of the flux surrounded by a $14'$ circular halo. We see no difference in the recovered beam shape in either case.

These results indicate that we do not resolve any of our sources and such effects are not present in our recovered holographic beam.

\subsection{Foreground contamination}
Another concern is Galactic or nearby point source contamination in the holographic measurements. To include the effects of the unknown primary beam of the tracking dish (in our case the 26 m), we simulate two noiseless models. The first is a a Gaussian 26 m beam, which therefore has no sidelobes. To include the effects of sidelobes of the tracking dish in a ``worst-case-scenario'' for a holographic measurement, we simulate an ``Exptan beam'' model which refers to the angular dependence of the theoretical illumination of a dipole feed in the aperture plane\cite{shaw2}. This model consists of a paraboloidal dish with the geometry of the 26 m illuminated by a Pathfinder dipole feed (essentially the dish version of the fiducial Pathfinder beam model). This is a ``worst-case scenario'' as its aperture (owing to the larger focal length of the 26 m) is over-illuminated leading to maximal sidelobes, and so does not reflect the reality of our 26 m setup.

\begin{figure}[ht] % not h only
	\centering	
	\includegraphics[width=0.95\textwidth]{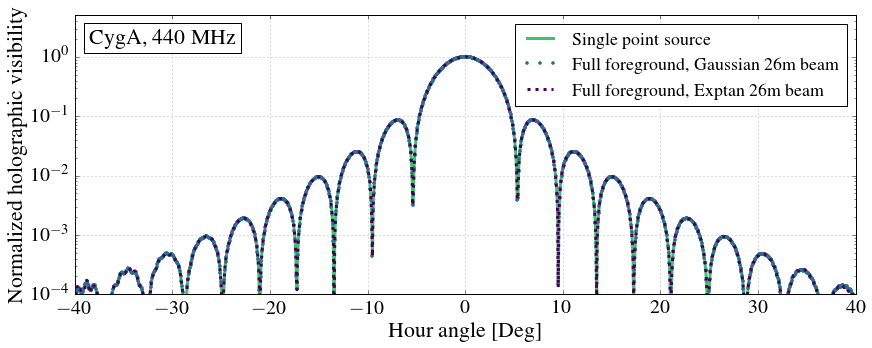}%
	\caption{The amplitudes of the Cygnus A beam slices for the single point source sky and full foreground models, in both cases with the Exptan 26 m beam, at 440 MHz. }
	\label{cygasim}
\end{figure}

Figure \ref{cygasim} shows the results for Cygnus A, which we have simulated at several different frequencies across the band and compare a sky with only one non-zero pixel to the full sky model. Cygnus A is the brightest source in the sky in the CHIME band and so shows the highest S/N sidelobe structure, but nonetheless has significant Galactic flux in its neighborhood. We see no deviation from the single point source slice in all cases.

As one descends in flux of the primary source, the distribution of point sources in the surroundings will change from Poissonian to Gaussian. We have also simulated the two less bright sources, Perseus B, and 3C295, but these are still among the brightest sources in the sky, and so we do not observe deviations between the single point source cases at any level of interest for our observations. To address the question of the minimum flux of primary source permitted for an unbiased measurement of the beam we require an accurate model of the 26 m, which we leave for a later work.

\section{Conclusion} \label{conclusion}

In this document, we have described and validated our technique of radio holography of bright astronomical point sources for obtaining high signal-to-noise (S/N) measurements, with good angular resolution, of the two-dimensional primary beams of the CHIME Pathfinder array across its frequency band. We have reported our progress in equipping the John A. Galt 26 m telescope with custom instrumentation for the purpose, and displayed the output of the method for a preliminary data set of 7 sources of minimal depth. 

It is clear from the data that more integration time is necessary for the low S/N sources to begin to measure the sidelobe structure seen in the best sources. From Figure \ref{full2dbeam} we see that the seven sources for which we have holography do not fully sample the two-dimensional structure in our model of the CHIME beams. The basic program is to use these holographic measurements to test and refine our beam models. Additionally, we plan to augment our holographic observations with additional sources which were not favorably located during the period these data were collected. However, our analysis of Section \ref{sec4} suggests that there is a minimum primary source flux at which one can expect to obtain a reasonable beam trace. In addition to the holography method presented here, we are pursing other methods of filling in the NS beam structure, notably with satellites\cite{hol2, sat2}, drones \cite{drone}, and pulsar holography. In its current form the data presented here serve as a basis for an understanding of the primary beam of a realistic cylindrical telescope array, such as CHIME and its Pathfinder.

\section{Acknowledgments}

We are grateful for the unflagging helpfulness of the DRAO staff and would like to particularly acknowledge Ev Sheehan, Rob Messing, Andrew Gray, and Kory Phillips for helpful discussions, use of the John A. Galt Telescope, and contributing valuable time in support of these observations.

We acknowledge support from the Canada Foundation for Innovation, the Natural Sciences and Engineering Research Council of Canada, the B.C. Knowledge Development Fund, le Cofinancement gouvernement du Qu\'ebec-FCI, the Ontario Research Fund, the CIfAR Cosmology and Gravity program, the Canada Research Chairs program, and the National Research Council of Canada.

\bibliography{hol_SPIEproc}
\bibliographystyle{spiebib}

%%%%%%%%%%%%%%%%%%%%%%%%%%%%%%%%%%%%%%%%%%%%%%%%%%%%%%%%%%%%%%%%%%%%%%%%
\end{document}